\documentclass{PoS}

\title{Nucleosynthetic Yields from ``Collapsars''}

\ShortTitle{Collapsars}

\author{
\speaker{Gabriel Rockefeller}$^{ab}$,
Christopher L. Fryer$^{ab}$, 
Patrick Young$^{ac}$, 
Michael Bennett$^{ad}$,
Steven Diehl$^{aeb}$,
Falk Herwig$^{adf}$,
Raphael Hirschi$^{adg}$,
Aimee Hungerford$^{ab}$,
Marco Pignatari$^{adh}$,
Georgios Magkotsios$^{ahc}$,
and Francis X. Timmes$^{ac}$ \\
\llap{$^a$}The NuGrid Collaboration\\
\llap{$^b$}Computational Physics and Methods (CCS-2), Los Alamos National Laboratory, Los Alamos, NM 87545, USA\\
\llap{$^c$}School of Earth and Space Exploration, Arizona State University, Tempe, AZ 85287, USA\\
\llap{$^d$}Astrophysics Group, Keele University, ST5 5BG, UK\\
\llap{$^e$}Theoretical Astrophysics (T-6), Los Alamos National Laboratory, Los Alamos, NM 87545, USA\\
\llap{$^f$}Dept. of Physics \& Astronomy, Victoria, BC, V8W 3P6, Canada\\
\llap{$^g$}IPMU, University of Tokyo, Kashiwa, Chiba 277-8582, Japan\\
\llap{$^h$}Joint Institute for Nuclear Astrophysics, University of Notre Dame, Notre Dame, IN 46556, USA\\
E-mail:\email{gaber@lanl.gov}
}

\abstract{The ``collapsar'' engine for gamma-ray bursts invokes as its
  energy source the failure of a normal supernova and the formation of
  a black hole.  Here we present the results of the first
  three-dimensional simulation of the collapse of a massive star down
  to a black hole, including the subsequent accretion and explosion.
  The explosion differs significantly from the axisymmetric scenario
  obtained in two-dimensional simulations; this has important
  consequences for the nucleosynthetic yields.  We compare the
  nucleosynthetic yields to those of hypernovae.  Calculating yields
  from three-dimensional explosions requires new strategies in
  post-process nucleosynthesis; we discuss NuGrid's plan for
  three-dimensional yields.}

\FullConference{10th Symposium on Nuclei in the Cosmos\\
		 July 27 - August 1 2008\\
		 Mackinac Island, Michigan, USA}

\begin{document}

\section{Explosions from the Collapse of Massive Stars}

Woosley~\cite{woo93} argued that a failed supernova could produce an
explosion after it collapsed down to a black hole, if the collapsing
star were rotating sufficiently rapidly.  This model, known as the
``collapsar'' model, has become the standard model for long-duration
gamma-ray bursts; it invokes a magnetic dynamo in the black hole
accretion disk~\cite{nar92,pop99} to power a relativistic jet and an
energetic stellar explosion, referred to as a hypernova.

\begin{figure}
\includegraphics[width=0.5\textwidth]{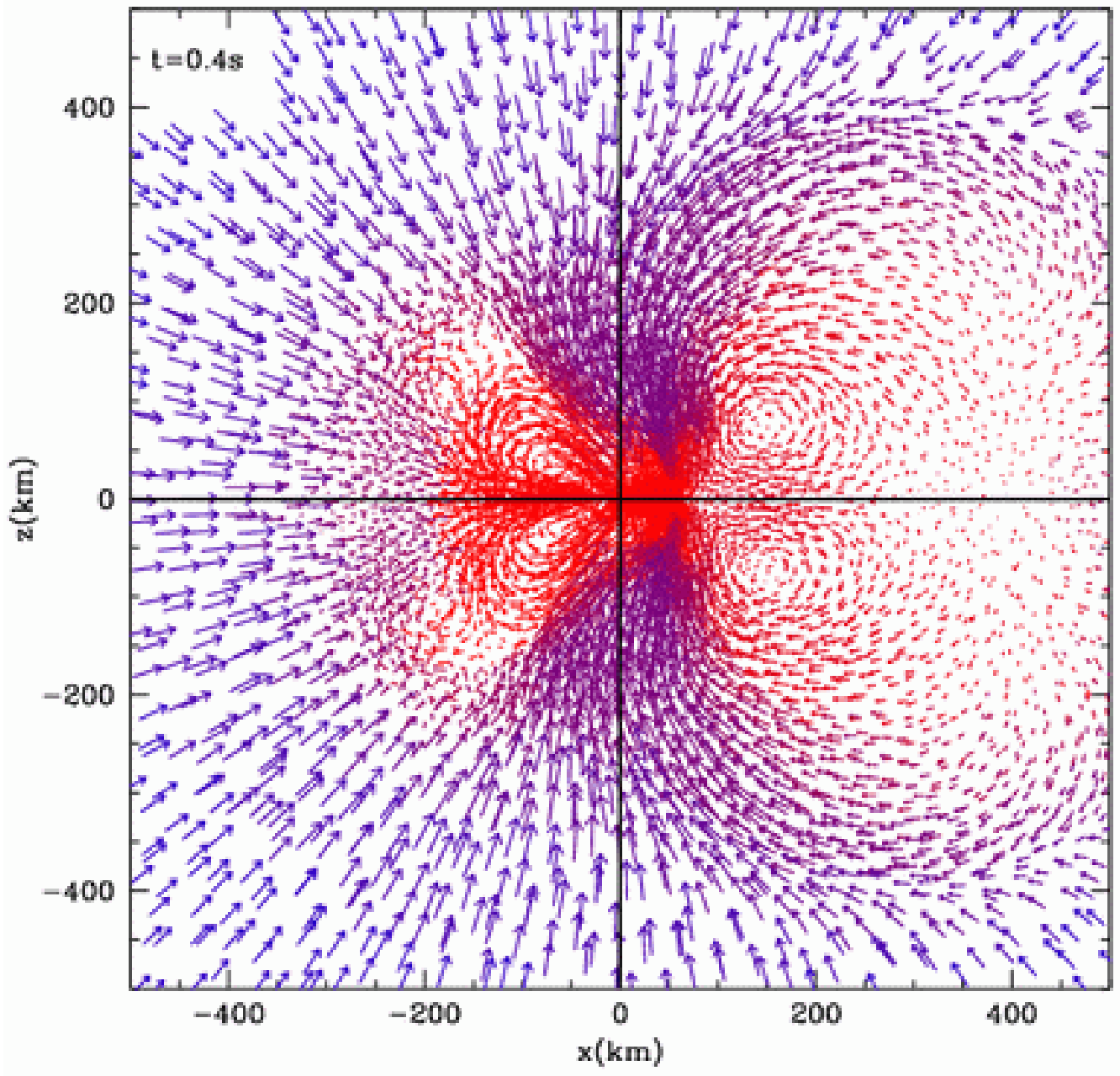}
\includegraphics[width=0.5\textwidth]{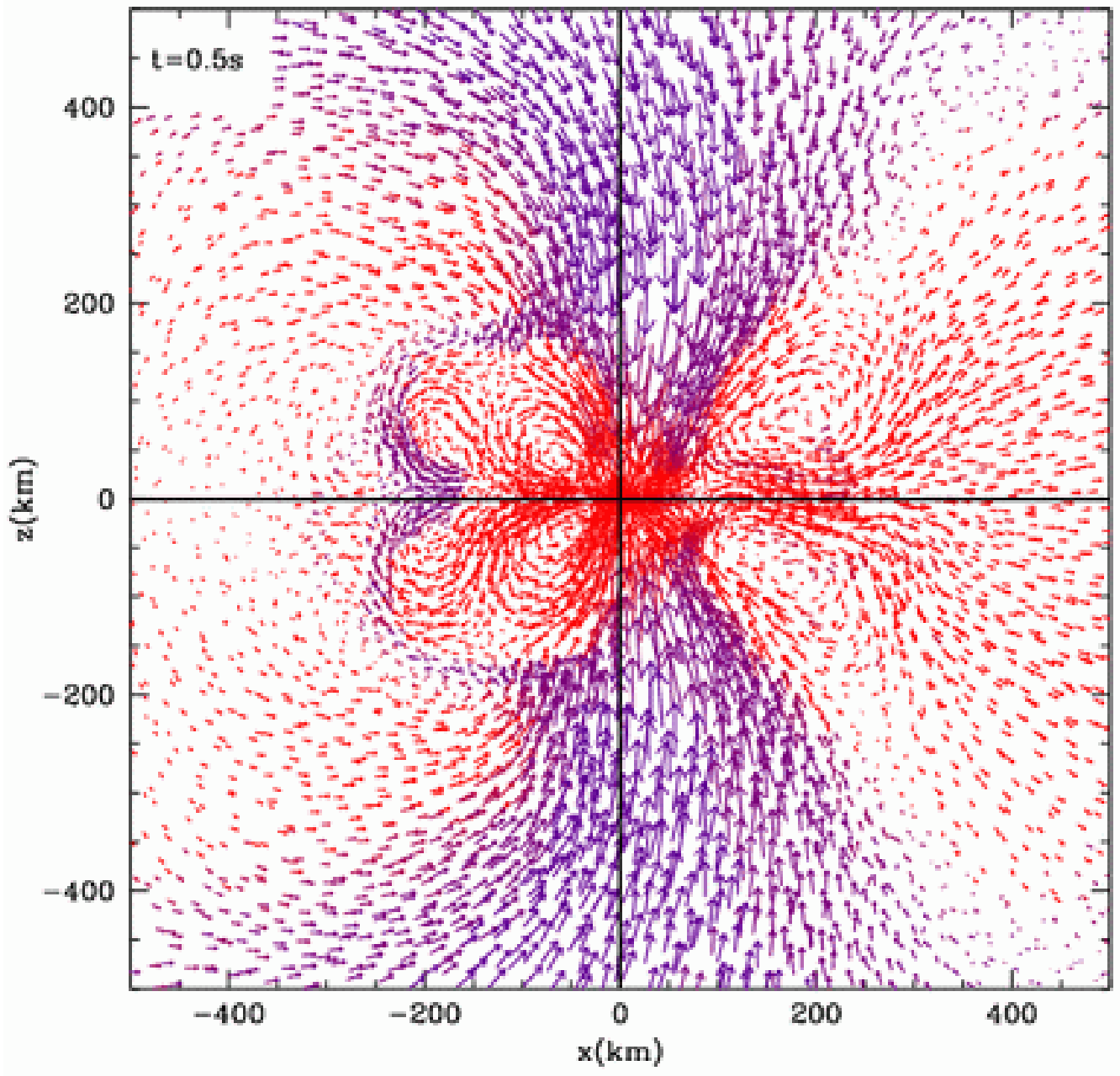}
\caption{Two slices through the x-z plane of a three-dimensional
  collapsar calculation.  Convection along the plane of rotation,
  driven by viscous heating from angular momentum transport,
  ultimately drives a strong explosion.  Although the explosion is
  hypernova-like in energy, it does not have the bipolar asymmetry
  that we have assumed for collapsars.}
\label{fig:expl}
\end{figure}

But what happens if the magnetic dynamo does not work?  Although
two-dimensional simulations showed that clean disks developed, which
were ideal for the formation of gamma-ray burst
jets~\cite{mac99,pro03}, theory predicted a much more chaotic system.
In the first three-dimensional simulations of the evolution of a
collapsar, we have found that instabilities in the disk lead to matter
ejection and hypernova-like explosions~\cite{rock08}.  These
explosions differ significantly from the jet-driven explosions that
Woosley first envisioned.  Figure~\ref{fig:expl} shows two slices (at
$0.4$\,s and $0.5$\,s after collapse) through the x-z plane from one
such simulation (where the z~axis is also the axis of rotation of the
star)~\cite{rock08}.  The progenitor of this explosion is a
60\,M$_\odot$ star evolved at zero metallicity, so it experienced low
mass loss; the final fate of such a star would be similar even at
metallicities as high as 1/100th solar.  The star collapses to a black
hole, and a disk forms from the fast-rotating silicon layer.  Viscous
heating drives convection and ultimately an explosion.

\begin{table}
\begin{tabular}{|l|c|c|c|c|}
\hline
Species & \multicolumn{4}{c|}{Yields (M$_\odot$)} \\
  & Rock1~\cite{rock08} & Tom50A~\cite{tom07} & Tom50B~\cite{tom07} & CL35~\cite{CL04} \\
\hline
\hline
$^{40}$Ca & 4.07E-02 & 1.87E-02 & 1.20E-02 & 1.70E-02 \\
$^{41}$Ca & 3.80E-07 & 1.72E-06 & 4.78E-07 & 2.07E-06 \\
$^{42}$Ca & 2.85E-06 & 2.29E-07 & 7.46E-08 & 1.54E-06 \\
$^{43}$Ca & 3.88E-09 & 7.31E-07 & 4.75E-07 & 1.56E-09 \\
$^{44}$Ca & 5.86E-08 & 6.08E-04 & 3.95E-04 & 1.03E-05 \\
$^{45}$Ca & 4.20E-09 & - & - & 7.22E-14 \\
$^{46}$Ca & 4.21E-08 & 2.66E-13 & 6.72E-14 & 3.96E-15 \\
$^{47}$Ca & 1.56E-09 & - & - & - \\
$^{48}$Ca & 1.35E-04 & 1.56E-13 & 6.72E-14 & 2.13E-19 \\
$^{45}$Sc & 5.04E-08 & 6.31E-08 & 3.69E-08 & 1.05E-07 \\
$^{44}$Ti & 9.23E-05 & 6.08E-04 & 3.95E-04 & 1.03E-05 \\
$^{46}$Ti & 8.86E-06 & 1.40E-05 & 9.07E-06 & 1.02E-06 \\
$^{47}$Ti & 4.89E-08 & 5.85E-05 & 3.80E-05 & 4.13E-08 \\
$^{48}$Ti & 9.09E-08 & 7.84E-04 & 5.09E-04 & 1.94E-04 \\
$^{49}$Ti & 9.86E-08 & 3.93E-06 & 2.55E-06 & 8.86E-06 \\
$^{50}$Ti & 2.93E-05 & 1.45E-12 & 5.46E-13 & 1.89E-13 \\
$^{50}$V & 6.42E-08 & 1.14E-12 & 2.83E-13 & 6.60E-12 \\
$^{51}$V & 5.06E-07 & 7.13E-05 & 4.63E-05 & 1.06E-05 \\
$^{50}$Cr & 3.12E-05 & 9.36E-06 & 6.06E-06 & 1.35E-05 \\
$^{52}$Cr & 1.78E-05 & 3.13E-03 & 2.03E-03 & 3.43E-03 \\
$^{53}$Cr & 1.21E-06 & 4.87E-05 & 3.15E-05 & 2.00E-04 \\
$^{54}$Cr & 6.28E-05 & 6.61E-12 & 1.72E-12 & 2.15E-10 \\
$^{55}$Mn & 1.42E-06 & 2.87E-05 & 1.86E-05 & 5.89E-04 \\
$^{54}$Fe & 4.35E-04 & 2.51E-05 & 1.63E-05 & 1.59E-03 \\
$^{56}$Fe & 7.15E-05 & 3.61E-01 & 2.34E-01 & 1.00E-01 \\
$^{57}$Fe & 3.47E-06 & 7.35E-03 & 4.77E-03 & 1.07E-03 \\
$^{58}$Fe & 1.68E-04 & 1.53E-11 & 4.52E-12 & 1.34E-10 \\
$^{60}$Fe & 1.04E-04 & 1.11E-12 & 3.86E-13 & 1.31E-21 \\
$^{59}$Co & 6.58E-07 & 1.34E-03 & 8.70E-04 & 7.58E-06 \\
$^{60}$Co & 4.64E-08 & - & - & 2.86E-16 \\
$^{56}$Ni & 4.74E-02 & 3.61E-01 & 2.34E-01 & 1.00E-01 \\
$^{57}$Ni & 4.55E-04 & 7.35E-03 & 4.77E-03 & 1.02E-03 \\
$^{58}$Ni & 2.32E-02 & 2.76E-03 & 1.79E-03 & 3.24E-04 \\
$^{60}$Ni & 2.42E-03 & 1.31E-02 & 8.49E-03 & 3.06E-04 \\
$^{61}$Ni & 4.43E-06 & 2.18E-04 & 1.42E-04 & 6.29E-06 \\
$^{62}$Ni & 9.44E-04 & 1.43E-04 & 9.31E-05 & 7.66E-06 \\
$^{64}$Ni & 4.93E-04 & 6.72E-12 & 1.67E-12 & 2.12E-15 \\

\hline

\end{tabular}
\caption{Yields from our explosion of a 60\,M$_\odot$ star
  compared to other low-metallicity calculations in the literature: two
  50\,M$_\odot$ stars~\cite{tom07} and a 35\,M$_\odot$ star~\cite{CL04}.
  The 50\,M$_\odot$ and 35\,M$_\odot$ stars differentiate between stable
  isotopes (that include the decay products into that isotope) whereas
  our explosion is the yield at the time of the explosion.}
\label{table:yield}
\end{table}

\section{Yields from Collapsars}

Without mass loss, 60\,M$_\odot$ stars are believed to collapse
directly to black holes~\cite{fry99}.  The possible outcomes of this
scenario cover a wide range of explosion energies and yields.  In the
limit where the progenitor star is non-rotating, such a collapse would
produce no explosion whatsoever.  At the other extreme, the collapse
of a rotating star could produce a hypernova~\cite{tom07}.  The
nucleosynthetic yields of these hypernovae are not too different from
the yields from standard nuclear yield studies~\cite{CL04}.  In this
section, we compare the yields from our three-dimensional simulation
to those of hypernovae and ``standard-yield'' explosions
(table~\ref{table:yield}).

To carry out our three-dimensional simulation, we used the SNSPH
code~\cite{FRW06} to model the evolution of over 2.5 million smooth
particle hydrodynamics (SPH) points covering the entire star.  For
this paper, we consider only the $\sim 250,000$ particles that burned
into iron-peak elements.  The Lagrangian nature of SPH allows us to
easily extract the density and temperature evolution of each particle
as a function of time.  We apply a post-process burning code to each
of these particles to obtain its nucleosynthetic yield using a
524-element network~\cite{young06}.

The results in table~\ref{table:yield} show that our explosion
produced considerably less $^{56}$Ni than either the hypernova
calculations~\cite{tom07} or the standard-yield
explosions~\cite{CL04}.  Our explosion is slightly less energetic than
the hypernova explosions, so the difference from the hypernova results
is not too surprising.  However, the fact that we do not produce as
much $^{56}$Ni as the standard nucleosynthetic yield calculations
highlights one of the major issues with these older yield results.
Models using piston-driven explosions artificially eject matter by
neglecting fallback~\cite{YF07}, which leads to overproduction of
$^{56}$Ni.  In addition, standard nucleosynthetic yield calculations
are driving explosions in massive stars (i.e., stars above
20\,M$_\odot$) that can only be expected in rapidly-rotating models
such as our collapsars or in hypernovae.  One of NuGrid's goals is to
determine the effect such assumptions have on the integrated yields.

There is a wealth of information in table~\ref{table:yield}, but here
we mention only a few other features.  The hypernova calculations
produce more $^{44}$Ti than our explosion; this is likely due to the
larger mass of high-entropy material produced in the more focused jet
explosion.  At high neutron number, the amounts of elements in the
hypernova and standard supernova explosions drop dramatically; this is
probably an artifact of the smaller nuclear networks used in those
calculations.  At the edges of networks, the yields are not very
reliable.  Our larger network produces more of these isotopes.

\section{Implications}

\begin{figure}
\center{\includegraphics[width=0.4\textwidth]{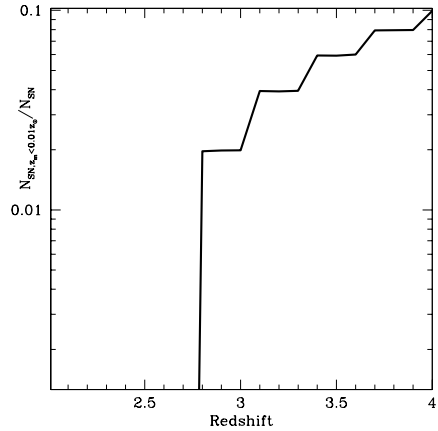}}
\caption{Fraction of stars with metallicities below 0.01Z$_\odot$ (a
  requirement for the direct collapse of 60\,M$_\odot$ stars to black
  holes) as a function of redshift.  The fraction drops off
  precipitously at a redshift of 2.8, so we do not expect the
  explosions simulated here to occur at lower redshifts.}
\label{fig:red}
\end{figure}

Will such explosions be observable?  Will their yields contribute
noticeably to the abundances of isotopes in the universe?  These
explosions will only occur in low-metallicity stars (i.e., stars with
metallicities less than 1/100th solar).  Figure~\ref{fig:red} shows
the fraction of stars formed that have such low metallicities as a
function of redshift.  Below a redshift of 2.8, very few stars have
sufficiently low metallicities to produce the explosions we observe.
This high redshift limit, coupled to our low simulated $^{56}$Ni
abundance (a large $^{56}$Ni abundance is very helpful in producing
bright light-curves), means that this type of explosion will not be
detected by any upcoming telescopes.  The lower yields from this
explosion, compared to earlier ``standard-yield'' simulations, imply
that massive stars have even less impact on the global abundance
pattern than has been previously thought.

The differences in the yields can be attributed to a number of
effects: different explosion mechanisms, different progenitors, and
different nuclear networks (interestingly, network uncertainties are
probably bigger than rate uncertainties at this point in time).  Our
NuGrid collaboration is studying each of these effects systematically
to reduce errors and to present a set of results including error bars
caused by current uncertainties.  We have also revised how we manage
data, introducing new data formats to deal will $>1$ million particle
simulations.


\begin{thebibliography}{99}
\bibitem{woo93} S. E. Woosley 1993, {\em Gamma-ray bursts from stellar
    mass accretion disks around black holes}, ApJ, {\bf 405}, 273
\bibitem{nar92} R. Narayan, B. Paczynski, T. Piran 1992, {\em
    Gamma-ray bursts as the death throes of massive binary stars},
  ApJ, {\bf 395}, L83
\bibitem{pop99} R. Popham, S. E. Woosley, C. Fryer 1999, {\em
    Hyperaccreting Black Holes and Gamma-Ray Bursts}, ApJ, {\bf 518},
  356
\bibitem{mac99} A. I. MacFadyen, S. E. Woosley 1999, {\em Collapsars:
    Gamma-Ray Bursts and Explosions in ``Failed Supernovae''}, ApJ,
  {\bf 524}, 262
\bibitem{pro03} D. Proga, A. I. MacFadyen, P. J. Armitage,
  M. C. Begelman 2003, {\em Axisymmetric Magnetohydrodynamic
    Simulations of the Collapsar Model for Gamma-Ray Bursts}, ApJ,
  {\bf 599}, L5
\bibitem{rock08} G. Rockefeller, C. L. Fryer, H. Li 2008, {\em
    Collapsars in Three Dimensions}, astro-ph/0608028
\bibitem{fry99} C. L. Fryer 1999, {\em Mass Limits for Black Hole
    Formation}, ApJ, {\bf 522}, 413
\bibitem{tom07} N. Tominaga, H. Umeda, K. Nomoto 2007, {\em Supernova
    Nucleosynthesis in Population III 13-15 M$_{\rm solar}$ Stars and
    Abundance Patterns of Extremely Metal-poor Stars}, ApJ, {\bf 660},
  516
\bibitem{CL04} A. Chieffi, M. Limongi 2004, {\em Explosive Yields of
    Massive Stars from Z=0 to Z=Z$_\odot$}, ApJ, {\bf 608}, 405
\bibitem{FRW06} C. L. Fryer, G. Rockefeller, M. S. Warren 2006, {\em
    SNSPH: A Parallel Three-dimensional Smoothed Particle Radiation
    Hydrodynamics Code}, ApJ, {\bf 643}, 292
\bibitem{young06} P. A. Young, C. L. Fryer, A. Hungerford, D. Arnett,
  G. Rockefeller, F. X. Timmes, B. Voit, C. Meakin, K. A. Eriksen
  2006, {\em Constraints on the Progenitor of Cassiopeia A}, ApJ, {\bf
    640}, 891
\bibitem{YF07} P. A. Young, C. L. Fryer 2007, {\em Uncertainties in
    Supernova Yields. I. One-Dimensional Explosions}, ApJ, {\bf 664},
  1033
\end{thebibliography}
\end{document}